\documentclass[preprint,12pt]{elsarticle}

\usepackage{amssymb}
\usepackage{amsmath}
\usepackage{pmat}
\usepackage{subfig}
\usepackage{caption}

\usepackage{graphicx}

\usepackage[retainorgcmds]{IEEEtrantools}

\usepackage[colorlinks,linkcolor=red,citecolor=red,urlcolor=red]{hyperref}

\usepackage{amsthm}
\newtheorem{thm}{Theorem}[section]
\newtheorem{lem}{Lemma}[section]
\newproof{pf}{Proof}
\newtheorem*{rmk-0}{Remark}
\newtheorem*{cor}{Corollary}
\newtheorem*{cor-1}{Corollary}
\newtheorem*{rmk-1}{Remark}
\newtheorem*{rmk-2}{Remark}

\makeatletter
\@addtoreset{equation}{section}
\makeatother
\renewcommand{\theequation}{\arabic{section}-\arabic{equation}}

\renewcommand{\theequationdis}{{\normalfont (\theequation)}}
\renewcommand{\theIEEEsubequationdis}{{\normalfont (\theIEEEsubequation)}}

\begin{document}

\begin{frontmatter}



\title{New Fundamental Formulas of Image Restoration in Spatial and Frequency Domains}


\author{Changcun Huang}


\begin{abstract}
Circular convolutions and the corresponding frequency domain formula are fundamentally important in image restoration; however, in this paper, we'll prove that the usual computing method of circular convolutions violates the physical meaning of blur producing. Especially for the image restoration algorithms in frequency domain, this violation will affect the restoration result. Relevant problems are proved rigorously and modified formulas are given in both spatial and frequency domains. Experiments are done to show the effects of new formulas. For clarity of proving, the one-dimensional case is dealt with first, which may be useful in one-dimensional signal processing.
\end{abstract}

\begin{keyword}
Image restoration; Circular convolution; Physical meaning; Signal processing.

\end{keyword}

\end{frontmatter}


\section{Introduction}
\begin{figure}[!t]
\captionsetup{justification=centering}
\centering
\includegraphics[width=2.045in]{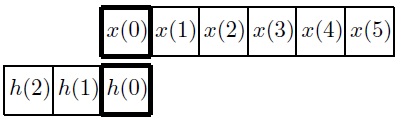}
\caption{A circular convolution.}
\label{Fig.1}
\end{figure}

Denote the circular convolution of one-dimensional signals as
\begin{IEEEeqnarray}{rCl}
y(n) = x(n) \circledast h(n),
\end{IEEEeqnarray}
where $h(n)$ is the convolution kernel of size $M \times 1$,  $x(n)$ and $y(n)$ are the original signal and convolved signal, respectively. As shown in Fig.\ref{Fig.1}, the computing method of circular convolutions is to reverse the element order of $h(n)$ to be $h'(n)$ and to make $h'(M-1)$ aligned with the processed element of $x(n)$, after which the weighted sum can be obtained.

The two-dimensional case is similar. In image restoration, the basic model (without considering noise) is $g(m,n) =  f(m,n) * h(m, n)$; if $DFT$ is used in frequency domain, the corresponding spatial domain operation will become a circular convolution
\begin{IEEEeqnarray}{rCl}
g(m, n) =  f(m, n) \circledast h(m, n),
\end{IEEEeqnarray}
where $g(m, n)$ is the degraded image, $f(m, n)$ is the original image, and $h(m, n)$ is the kernel as well as the point spread function ($PSF$). The computing of (1-2) is done in two dimensions separately by the method of (1-1). The goal of image restoration is to reconstruct the original image $f(m, n)$ from the degraded observation image $g(m, n)$ by getting the solution of $f(m, n)$. By $DFT$, (1-2) has its equivalent form in frequency domain as
\begin{IEEEeqnarray}{rCl}
\mathcal{G}(u,v) = \mathcal{F}(u,v)\mathcal{H}(u,v),
\end{IEEEeqnarray}
which is very useful and is a fundamental formula in image restoration \cite{[2]}. For example, IBD \cite{[1],[3]} is a classic algorithm in blind image restoration and its main operations are in frequency domain, which is a direct use of (1-3). The methods of Wiener filter \cite{[4]} and constrained least square estimation \cite{[5]} are also related to (1-3).

In this paper, we'll prove that circular convolutions in terms of both spatial and frequency domains violate the physical meaning of blur producing in image restoration; and new formulas will be given.

\section{Physical meaning of the $PSF$ convolution}
\subsection{Symmetric-property problem}
The sources of image blur are mainly classified into three categories \cite{[6]}: defocus blur, atmospheric-turbulence blur and motion blur. All of their $PSF$ can be modeled by masks (also called filters or kernels). The weight values of the $PSF$ mask determine the degradation type.

\begin{figure}[!t]
\captionsetup{justification=centering}
\centering
\subfloat[An example of $PSF$ convolutions.]{\includegraphics[width=2.3in]{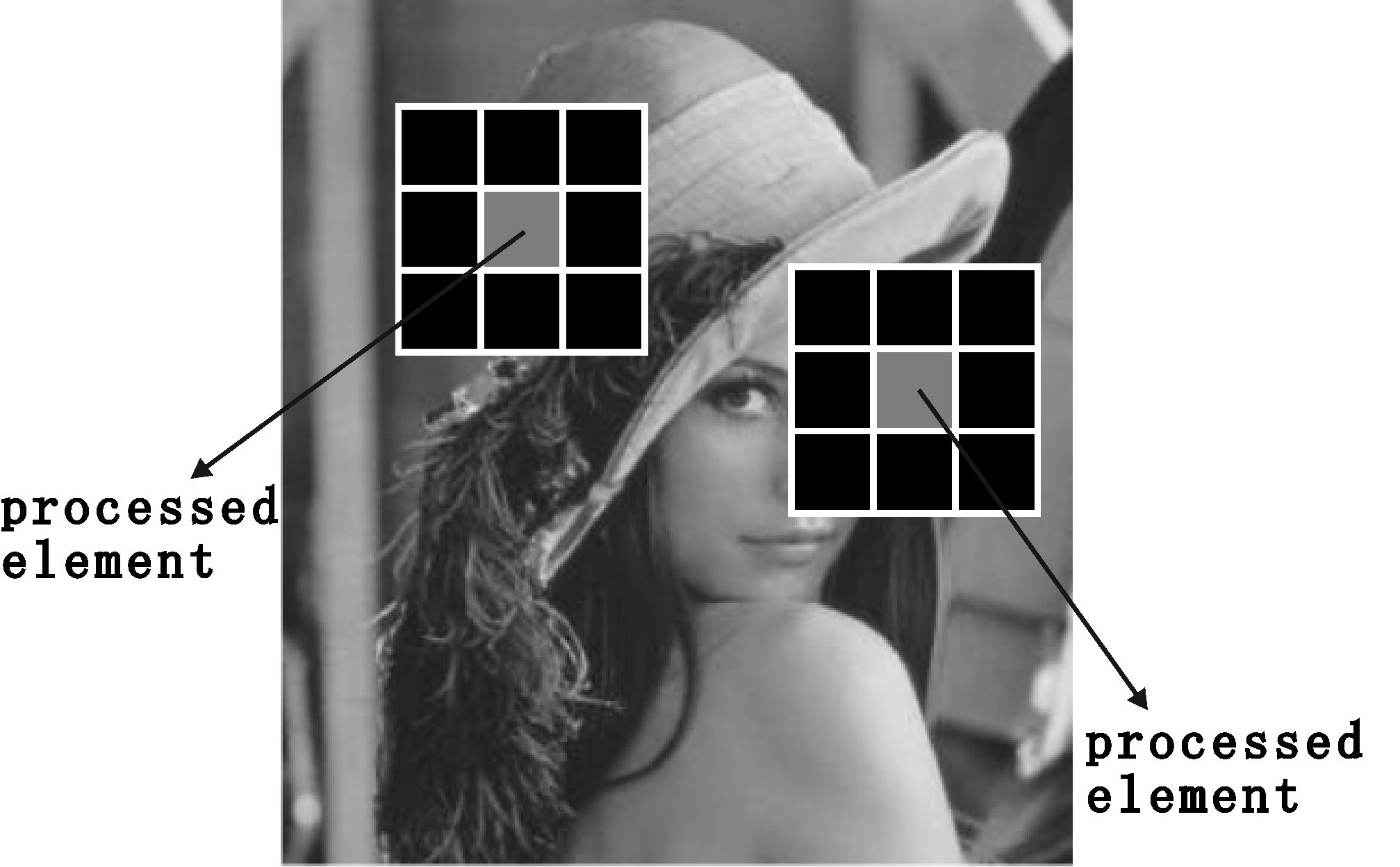}} \quad \quad
\subfloat[A $PSF$ mask of horizontal motion blur.]{\includegraphics[width=1.45in]{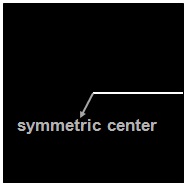}}
\caption{Symmetric property of $PSF$ convolutions.}
\label{Fig.2}
\end{figure}

The symmetric property of the $PSF$ convolution means that the image pixel to be processed should be aligned with the symmetric center of the $PSF$ mask. Fig.\ref{Fig.2}(a) gives an example. The big white square is a $PSF$ mask of size $3 \times 3$. The gray-filled small square in the big square is the symmetric center of the $PSF$ mask whose size equals one pixel of the image (in the sense of principles). The location of the gray-filled square is exactly the processed image pixel. This symmetric property has clear physical meaning in producing of defocus blur and atmospheric-turbulence blur and also is required in the blind deconvolution of motion blur.

The $PSF$ of defocus blur is \cite{[6]}：
\begin{equation*}
h(x,y) =
\begin{cases}
\frac{1}{\pi R^2}, & \text{if} \quad \sqrt{x^2 + y^2} \le R
\\
0, & \text{otherwise}
\end{cases},
\end{equation*}
which is centrosymmetric and its symmetric center is $(0, 0)$.
The pixel to be convolved should be aligned with this center and the new pixel value is the weighted sum of neighborhood pixels; and so is the case of atmospheric-turbulence blur whose $PSF$ is $h(x,y)=Ke^{-\frac{x^2 + y^2}{2\sigma^2}}$ \cite{[6]}.

Motion blur is due to a relative motion between the scene to be imaged and the camera during exposure. Despite non-symmetric in weight values, motion blur $PSF$ can be modeled by a square mask with nonzero weight values in the motion direction. The $PSF$ of horizontal motion blur with uniform velocity is \cite{[7]}:
\begin{IEEEeqnarray*}{rCl}
h(x,y)=\frac{1}{\alpha_0}rect(\frac{x}{\alpha_0}-\frac{1}{2})\delta(y),
\end{IEEEeqnarray*}
where $rect(x) = 1$ for $|x|\le\frac{1}{2}$ and $rect(x) = 0$ otherwise, and $\delta(y)$ is the Dirac delta function. As shown in Fig.\ref{Fig.2}(b), it's a $PSF$ mask of horizontal motion blur whose weight values are nonzero in the horizontal white line, by which a motion blur image in horizontal direction could be produced. In the perspective of mask models, the convolution process of motion blur is the same as defocus or atmospheric-turbulence blur. This symmetric mask model is useful in blind deconvolution of motion blur images, since the motion direction is unknown and all possible directions should be taken into consideration.

\subsection{Reverse-order problem}
As mentioned above, the computing of circular convolutions should reverse the element order of a $PSF$; however, if the $PSF$ is not centrosymmetric, such as the motion blur $PSF$ of Fig.\ref{Fig.2}(b), the reverse operation will change the weight values of blur producing.

The topics of this paper are all about the two problems discussed above.

\section{The one-dimensional case}
We first discuss the one-dimensional case in order to clarify the proof; meanwhile, the results may be useful in one-dimensional signal processing.

Let $x(n)$ and $y(n)$ be vectors of size $N \times 1$. $h(n)$ is the kernel for $n=0,1,\cdots, M-1$, where $M$ is an odd integer with $M<N$. $h_e(n)$ is the extended form of $h(n)$ defined as \cite{[5]}
\begin{equation*}
h_e(n) =
\begin{cases}
h(n) & \text{for } 0<n\le M-1
\\
0 & \text{for } M \le n \le N-1
\end{cases}.
\end{equation*}

If we denote the matrix form of (1-1) as
\begin{IEEEeqnarray}{rCl}
y = Hx,
\end{IEEEeqnarray}
then
\begin{IEEEeqnarray}{rCl}
H=\begin{pmatrix}
h_e(0) & h_e(N-1) & \cdots & h_e(1)\\
h_e(1) & h_e(0) & \cdots & h_e(2)\\
\vdots & &\ddots & \vdots\\
h_e(N-1) & h_e(N-2) & \cdots & h_e(0)\\
\end{pmatrix},
\end{IEEEeqnarray}
which is called convolution matrix and is a circulant matrix \cite{[8]}.

The frequency domain formula of (1-1) is
\begin{IEEEeqnarray}{rCl}
\mathcal{Y}(k) = \mathcal{X}(k)\mathcal{H}(k),
\end{IEEEeqnarray}
where $\mathcal{X}(k)$, $\mathcal{H}(k)$, and $\mathcal{Y}(k)$ are the $DFT$ of $x(n)$, $h_e(n)$ and $y(n)$, respectively.

\begin{figure}[!t]
\captionsetup{justification=centering}
\centering
\includegraphics[width=2.045in]{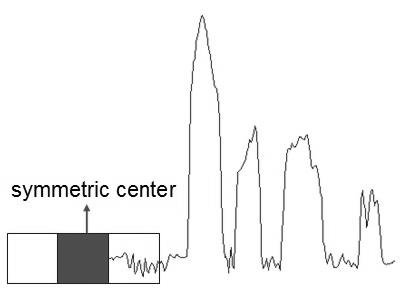}
\caption{Symmetric property of one-dimensional case.}
\label{Fig.3}
\end{figure}

The symmetric property of the one-dimensional case is similar to that of two dimensions. As in Fig.\ref{Fig.3}, the time varying curve is a signal and the rectangle is a kernel with its symmetric center gray filled. It shows that the signal element to be processed is aligned with the symmetric center of the kernel, which is the one-dimensional symmetric property.

\subsection{Problems of existing formulas}

\begin{thm}
Denote
\begin{IEEEeqnarray}{rCl}
h'(n)=[h(M-1) \quad h(M-2) \quad \cdots\ \quad h(0)]^T.
\end{IEEEeqnarray}
The usual computing method of circular convolution {\rm(3-1)} doesn't satisfy the symmetric property; the $M - 1$th element of $h'(n)$ instead of its symmetric center is aligned with the processed signal element. Also, the order of weight values is not as $h(n)$, but as $h'(n)$ of {\rm(3-4)}, i.e., the reverse of $h(n)$.
\end{thm}
\begin{pf}
The conclusions are obvious which can be seen from Fig.\ref{Fig.1} and the proof is trivial; we just provide some details more formally and introduce several terms for use. First, describe the ``convolution meaning" of matrix $H$ in (3-1) and (3-2). Each row of $H$ in (3-2) corresponds to the convolution of each element of $x(n)$. The convolution structure, i.e., the way how to sum the values of $x(n)$ in a neighborhood, is determined by the first row of matrix $H$, which is the same for all elements of $x(n)$ by the property of circulant matrices. Different rows of $H$ are only different positions of the kernel with the convolution structure unchanged. Therefore, analysing the first row of $H$ and the first convolved signal element $y(0)$ is enough.

The first processed signal element is $x(0)$ and the convolved result is $y(0)$. By (3-1) and (3-2), we have
\begin{IEEEeqnarray*}{rCl}
y(0) = [h_e(0) \quad h_e(N-1) \quad \cdots \quad h_e(1)]
\nonumber \\*
\cdot\lbrack{x(0) \quad x(1) \quad \cdots \quad x(N - 1)}\rbrack^T.
\end{IEEEeqnarray*}
Rearrange the element order of the two multiplying vectors and make the sum $y(0)$ unchanged simultaneously in such a way
\begin{IEEEeqnarray*}{rCl}
y(0) = [h_e(N-1) \quad \cdots \quad h_e(1) \quad h_e(0)]
\nonumber \\
\cdot\lbrack{x(1) \quad \cdots \quad x(N - 1) \quad x(0)}\rbrack^T.
\end{IEEEeqnarray*}
By definition, $h_e(n)$ is zero when $n \ge M$ and equals $h(n)$ otherwise, which follows
\begin{IEEEeqnarray}{rCl}
\IEEEeqnarraymulticol{3}{l}{
y(0) = [h(M-1) \quad \cdots \quad h(1) \quad h(0)]}
\nonumber \\ \qquad
\cdot\lbrack{x(N-(M-1)) \quad \cdots x(N - 1) \quad x(0)}\rbrack^T.
\end{IEEEeqnarray}

Fig.\ref{Fig.1} is an example of (3-5) when $N = 6$ and $M = 3$. We can see that the kernel is $h'(n)$ instead of $h(n)$; and the symmetric center $h(1)$ of $h'(n)$ is not aligned with the processed signal element $x(0)$, while the third element $h'(2)$ is aligned instead. The general case is similar.

\end{pf}

\begin{rmk-0}
The problem of the symmetric property or reverse order may also exist in one-dimensional case, if the convolution process has symmetric physical meaning or the kernel is not centrosymmetric.
\end{rmk-0}

By the above discussions, the following conclusion in frequency domain is natural.
\begin{cor}
If the frequency domain formula of a circular convolution is {\rm(3-3)}, then the corresponding time domain convolution does not satisfy the symmetric property, and the order of weight values is as $h'(n)$ of {\rm(3-4)} instead of $h(n)$.
\end{cor}
\begin{pf}
The corresponding time domain formula of (3-3) is (3-1), by Theorem 3.1, this corollary holds.
\end{pf}

\subsection{Modified formulas}
\begin{lem}
The symmetric center of $h'(n)$ of {\rm{(3-4)}} can be made aligned with the processed signal element by moving the elements of matrix $H$ of {\rm(3-2)} in terms of
\begin{IEEEeqnarray}{rCl}
H_t=H\alpha^{t},
\end{IEEEeqnarray}
where
\begin{IEEEeqnarray}{rCl}
\alpha=\begin{pmatrix}
0 & 1 & 0  \cdots & 0 \cr\-
0 & 0 & 1  \cdots & 0 \cr\-
\vdots &  &  \ddots & \vdots \cr\-
1 & 0 & 0  \cdots & 0 \cr
\end{pmatrix}
\end{IEEEeqnarray}
is a $N \times N$ permutation matrix {\rm\cite{[8]}} and
\begin{IEEEeqnarray}{rCl}
t=\frac{M-1}{2}.
\end{IEEEeqnarray}
\end{lem}
\begin{pf}
As stated in literature \cite{[8]}, when multiplying a vector such as $h_e^T$ (size $1 \times N$) by permutation matrix $\alpha$, it means that all the elements of $h_e^T$ move one position to the right and wrap around, which is actually a forward shift permutation
\begin{IEEEeqnarray*}{rCl}
h_e^T\alpha = \sigma(h_e^T) =  [h_e(N-1) \quad h_e(0) \quad \cdots \quad h_e(N-2)],
\end{IEEEeqnarray*}
where $\sigma$ is a permutation operator. So multiplying $h_e^T$ by $\alpha^t$ means to do this operation by $t$ times. Obviously, moving the elements is equivalent to moving the position of a kernel.

Write matrix $H$ as
\begin{IEEEeqnarray*}{rCl}
H = [H_1 \quad H_2 \quad \cdots \quad H_N]^T.
\end{IEEEeqnarray*}
where $H_i$ is the $i$th row of $H$, by which (3-6) can be expressed as
\begin{IEEEeqnarray}{rCl}
H_t = [H_1\alpha ^t \quad H_2\alpha ^t \quad \cdots \quad H_N\alpha ^t]^T.
\end{IEEEeqnarray}

From (3-9), in each row of matrix $H$, if we move each element $t$ positions to the right and wrap around, the result is $H_t$. By Theorem 3.1, when the convolution matrix is $H$, $h'(M-1)$ is aligned with the processed signal element. If we move the kernel right by
\begin{IEEEeqnarray*}{rCl}
t = M - 1 - \frac{M - 1}{2} = \frac{M - 1}{2}
\end{IEEEeqnarray*}
steps, the symmetric property will be satisfied.

\begin{figure}[!t]
\captionsetup{justification=centering}
\centering
\subfloat[Convolution with matrix $H$.]{\includegraphics[width=2.045in]{3_a}} \quad \quad
\subfloat[Convolution with matrix $H_t$.]{\includegraphics[width=2.1in]{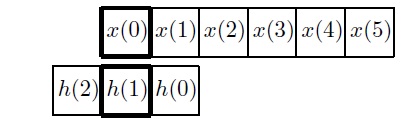}}
\caption{Effect of moving a kernel in Lemma 3.1.}
\label{Fig.4}
\end{figure}

Fig.\ref{Fig.4} shows an example. For convenance of descriptions and comparisons, Fig.\ref{Fig.1} is repeated in Fig.\ref{Fig.4}(a), which is the convolution by matrix $H$. After the permutation operation of matrix $\alpha$, Fig.\ref{Fig.4}(b) of matrix $H_t$ satisfies the symmetric property.

Note that this operation doesn't change the element order of the kernel, so the kernel is still (3-4) instead of the original $h(n)$.
\end{pf}

\begin{thm}
In {\rm(3-1)}, change $h(n)$ into $h'(n)$ of {\rm(3-4)}, and construct a convolution matrix $H'_t$ via Lemma 3.1, then
\begin{IEEEeqnarray}{rCl}
y = H_t'x
\end{IEEEeqnarray}
satisfies the symmetric property and the convolution structure is the original $h(n)$.
\end{thm}
\begin{pf}
The proof is trivial by the computing method of circular convolutions and Lemma 3.1. When constructing the convolution matrix by $h'(n)$, because the convolution operation should reverse the element order of the kernel, the reversed $h'(n)$ is just $h(n)$. The effect of $H_t'$ by Lemma 3.1 realizes the symmetric property. This completes the proof.
\end{pf}

\begin{lem}
The frequency domain formula of
\begin{IEEEeqnarray}{rCl}
y = H_tx
\end{IEEEeqnarray}
is
\begin{IEEEeqnarray}{rCl}
\mathcal{Y}(k) = \mathcal{X}(k)\mathcal{H}(k)e^{\frac{i2\pi}{N}kt},
\end{IEEEeqnarray}
where $H_t$ is defined in {\rm(3-6)} of Lemma {\rm{3.1}} and $t$ is as in {\rm(3-8)}.
\end{lem}
\begin{pf}
Also the proof begins with an example of the first convolution place in Fig.\ref{Fig.4}(a). Fig.\ref{Fig.4}(a) is the convolution by matrix $H$ of (3-1); for clarity, in this proof, denote the convolved signal of $H$ by $y_0(n)$. In Fig.\ref{Fig.4}(a), if $x(5)$ instead of $x(0)$ is considered as the processed element, the symmetric property would be satisfied; so the first element of $y_0(n)$ can be regarded as the convolution result of $x(5)$, which is previous $x(0)$ (in the sense of zigzag order). Similarly, each element of $y_0(n)$ for other $n's$ can also be considered as the convolution result of previous element.

Because $y(n)$ of (3-11) is obtained by convolutions satisfying the symmetric property (according to Lemma 3.1), $y_0(n)$ is in fact a signal obtained by shifting the elements of $y(n)$ right by one step. According to the circular shift property of $DFT$, we have $\mathcal{Y}_0(k) = \mathcal{Y}(k)e^{\frac{-2\pi}{N}k}$, i.e.,
\begin{IEEEeqnarray}{rCl}
\mathcal{Y}(k) = \mathcal{Y}_0(k)e^{\frac{2\pi}{N}k},
\end{IEEEeqnarray}
where $\mathcal{Y}(k)$ and $\mathcal{Y}_0(k)$ are the $DFT$ of $y(n)$ and $y_0(n)$, respectively. Since $\mathcal{Y}_0(k) = \mathcal{X}(k)\mathcal{H}(k)$, it follows that
\begin{IEEEeqnarray}{rCl}
\mathcal{Y}(k) = \mathcal{X}(k)\mathcal{H}(k)e^{\frac{2\pi}{N}k}.
\end{IEEEeqnarray}

(3-14) is the frequency domain formula of Fig.\ref{Fig.4}(b) when the kernel length is 3; the general case is the formula of (3-12), which can be similarly obtained by the method above.

\end{pf}

Based on Lemma 3.2, we present the modified form of (3-3) in frequency domain.
\begin{thm}
If the frequency domain formula of {\rm{(3-3)}} is modified to
\begin{IEEEeqnarray}{rCl}
\mathcal{Y}(k) = \mathcal{X}(k)\mathcal{H}'(k)e^{\frac{i2\pi}{N}kt},
\end{IEEEeqnarray}
where $\mathcal{H}'(k)$ is the $DFT$ of $h_e'(n)$, which is the extended form of $h'(n)$ of {\rm(3-4)}, then its corresponding time domain convolution satisfies the symmetric property and the kernel is the original $h(n)$.
\end{thm}
\begin{pf}
\par
(3-12) is the frequency domain formula of matrix $H_t$ that satisfies the symmetric property as mentioned in Lemma 3.2; however, the kernel or the convolution structure of $H_t$ is not $h(n)$ but the reverse order of $h(n)$. Noting that $\mathcal{H}(k)$ of (3-12) is corresponding to the convolution of $h'(n)$, while $h'(n)$ is the reverse order of $h(n)$, if we modify $\mathcal{H}(k)$ to be the $DFT$ of $h'_{e}(n)$, then its corresponding time domain kernel will be the reverse order of $h'(n)$, which is just $h(n)$. Therefore, (3-15) is the final result.
\end{pf}

\begin{rmk-1}
\rm{(3-10)} and \rm{(3-15)} are the modified formulas of circular convolutions in time domain and frequency domain, respectively. They are corresponding to each other in different domains.
\end{rmk-1}

\subsection{Summary}
In this section, the one-dimensional case was discussed. Next we'll generalize it to the two-dimensional image restoration.

\section{The two-dimensional case}
Let's give more detailed descriptions of the notations related with image restoration in Section 1. Denote the original image and degraded image by $f(m, n)$ and $g(m, n)$, respectively, for $0\le m \le M-1$ and $ 0 \le n \le N-1$. $h(m, n)$ is the $PSF$ for $ 0 \le m \le J-1$ and $0 \le n \le K-1$, where $J$ and $K$ are odd integers. $h_e(m, n)$ is the extended form of $h(m, n)$:
\begin{IEEEeqnarray*}{rCl}
h_e(m, n) = \begin{cases}
h(m, n) & \text{for } 0 \le m \le J-1  \text{ and } 0\le n \le K-1
\\
0 & \text{for } J \le m \le M-1 \text{ or } K \le n \le N-1
\nonumber
\end{cases}.
\end{IEEEeqnarray*}

$\mathcal{F}(u,v)$, $\mathcal{G}(u,v)$ and $\mathcal{H}(u,v)$ are the $DFT$ of $f(m, n)$, $g(m, n)$ and $h_e(m, n)$, respectively.

$\vec{g}$ is the vector-matrix form of matrix $g$, which represents the two-dimensional matrix by zigzag order. $\vec{f}$ is similar.

Rewrite the spatial domain and frequency domain formulas of image restoration here (with noise ignored):
\begin{IEEEeqnarray}{rCl}
g(m,n) & = & f(m,n) \circledast h(m,n)
\end{IEEEeqnarray}
and
\begin{IEEEeqnarray}{rCl}
\mathcal{G}(u,v) = \mathcal{F}(u,v)\mathcal{H}(u,v).
\end{IEEEeqnarray}
The matrix form of (4-1) is
\begin{IEEEeqnarray}{rCl}
\vec{g} = H\vec{f},
\end{IEEEeqnarray}
where $H$ is a block circulant matrix \cite{[8]} as
\begin{IEEEeqnarray}{rCl}
H=\begin{pmat}({||})
H_0 & H_{M-1}  \cdots & H_1 \cr\-
H_1 & H_0  \cdots & H_2 \cr\-
\vdots & \ddots & \vdots \cr\-
H_{M-1} & H_{M-2}  \cdots & H_0 \cr
\end{pmat}
\end{IEEEeqnarray}
with $H_i(0 \le i \le M-1)$ being
\begin{IEEEeqnarray}{rCl}
H_i = \begin{pmatrix}
h_e(i,0) & h_e(i,N-1)\cdots & h_e(i,1)\\
h_e(i,1) & h_e(i,0)\cdots & h_e(i,2)\\
\vdots &\ddots & \vdots\\
h_e(i,N-1) & h_e(i,N-2)\cdots & h_e(i,0)\\
\end{pmatrix}.
\nonumber \\*
\end{IEEEeqnarray}

\subsection{Problems of existing formulas}

\begin{figure}[!t]
\captionsetup{justification=centering}
\centering
\includegraphics[width=2.045in]{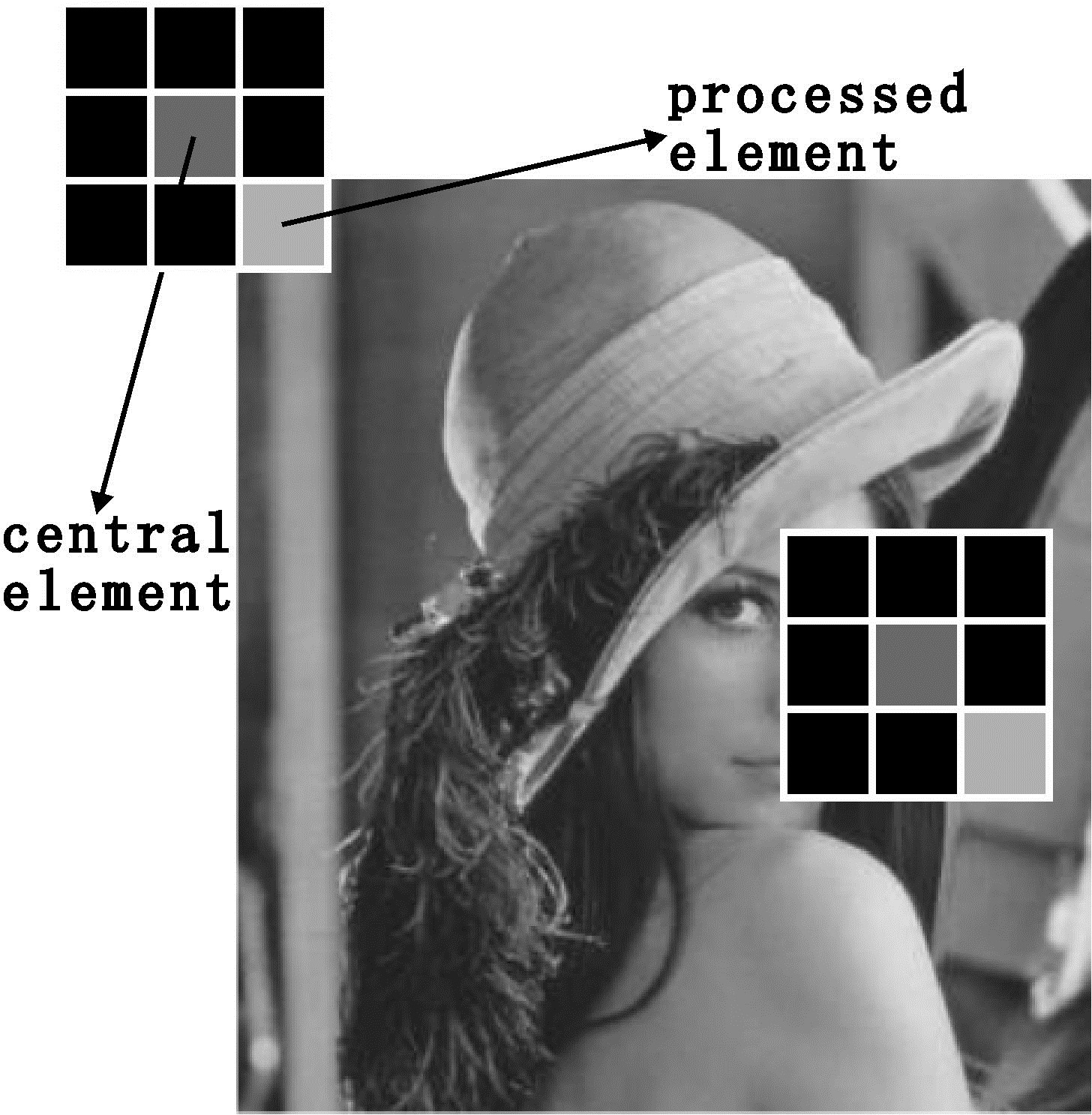}
\caption{An example of Theorem 4.1.}
\label{Fig.5}
\end{figure}

\begin{thm}
The circular convolution of {\rm{(4-3)}} doesn't satisfy the symmetric property and its $PSF$ is changed into
\begin{IEEEeqnarray}{rCl}
h'(m, n) = h(\tau(m), \tau'(n)),
\end{IEEEeqnarray}
where $\tau(m) = M-1-m$ and $\tau'(n) = N - 1 - n$ are both reverse operations. Element $h'(M - 1, N - 1)$ of the $PSF$ mask is aligned with the processed image pixel, instead of the symmetric center $h'(\frac{M - 1}{2}, \frac{N - 1}{2})$.
\end{thm}
\begin{pf}
As long as we notice the convolution meaning of block circulant matrix $H$ of (4-4), the two-dimensional generalization of Theorem 3.1 is trivial. Write the expanded form of (4-3) as:
\begin{IEEEeqnarray}{rCl}
\pmatset{0}{4pt}
\begin{pmat}({})
\vec{g_0} \cr\-
\vec{g_1} \cr\-
\vdots \cr\-
\vec{g_{M-1}} \cr
\end{pmat}=
\begin{pmat}({||})
H_0 & H_{M-1} \cdots & H_1 \cr\-
H_1 & H_0 \cdots & H_2 \cr\-
\vdots & \ddots & \vdots \cr\-
H_{M-1} & H_{M-2} \cdots & H_0 \cr
\end{pmat}
\begin{pmat}({})
\vec{f_0} \cr\-
\vec{f_1} \cr\-
\vdots \cr\-
\vec{f_{M-1}} \cr
\end{pmat},
\nonumber \\*
\end{IEEEeqnarray}
where $\vec{g_i}$ is the $i$th row of a degraded image and $\vec{f_i}$ is the $i$th row of an original image. The convolution structure of (4-7) is: The block operations of $H$ correspond to the row convolutions, while the operations of each block-element matrix are related to the column convolutions.

For example, if the $PSF$ mask size is $3 \times 3$, the first block of $\vec{g}$ (i.e., the first row of image $g$) is
\begin{IEEEeqnarray}{rCl}
\vec{g_0} = [H_{2} \quad H_1 \quad H_0][\vec{f_{M-2}} \quad \vec{f_{M-1}} \quad
\vec{f_0}]^T,
\end{IEEEeqnarray}
which means that the rows of $M-2$, $M-1$, $0$ of matrix $f$ participate in the convolution of the first row of matrix $g$.

The case of column convolutions can be found in the matrix operations of each block element (such as $H_0$ in (4-4)), which is actually the one-dimensional case. For example, in (4-8), $H_0\vec{f_0}$ deals with the column dimension of row 0.

Only considering the blocks of $H$, using Theorem 3.1 can prove the row-dimensional case. The case of column-dimension can be proved by applying Theorem 3.1 to each $H_i$ of (4-5). Fig.\ref{Fig.5} is a visualization of this theorem.
\end{pf}

The corresponding conclusion in frequency domain is as follows.
\begin{cor-1}
If the frequency domain formula of a circular convolution is {\rm(4-2)}, then the corresponding spatial domain operations do not satisfy the symmetric property, and the structure of weight values is as $h'(m, n)$ of {\rm(4-6)} instead of $h(m, n)$.
\end{cor-1}
\begin{pf}
The proof is similar to the one-dimensional case.
\end{pf}

\subsection{Modified formulas}
\begin{lem}
The symmetric center of $h'(m, n)$ in {\rm(4-6)} can be aligned with the processed image pixel, if the block circulant matrix $H$ of {\rm{(4-4)}} is changed into
\begin{IEEEeqnarray}{rCl}
H_t = H\alpha_1^sC(\alpha_2^t),
\end{IEEEeqnarray}
where both $\alpha_1^s$ and $C(\alpha_2^t)$ are permutation operations. $\alpha_1$ is actually a permutation matrix of size $M \times M$ as {\rm(3-7)}
\begin{IEEEeqnarray}{rCl}
\pmatset{0}{4pt}
\alpha_1=\begin{pmat}({|||})
0 & 1 & 0  \cdots & 0 \cr\-
0 & 0 & 1 \cdots & 0 \cr\-
\vdots &   & \ddots & \vdots \cr\-
1 & 0 & 0  \cdots & 0 \cr
\end{pmat},
\end{IEEEeqnarray}
which is used for permutation operations of rows; we designate it here as a block matrix since $H\alpha_1^s$ only refers to the block-matrix multiplication. $C(\alpha_2^t)$ is related to the permutation operations of columns and is a diagonal block matrix of size $M \times M$:
\begin{IEEEeqnarray}{rCl}
\pmatset{0}{4pt}
C(\alpha_2^t) =  \begin{pmat}({|||})
\alpha_2^t & 0 & 0  \cdots & 0 \cr\-
0 & \alpha_2^t & 0  \cdots & 0 \cr\-
\vdots &  &  \ddots & \vdots \cr\-
0 & 0 & 0  \cdots & \alpha_2^t \cr
\end{pmat},
\end{IEEEeqnarray}
where $\alpha_2$ is a permutation matrix of size $N \times N$. $s$ and $t$ are the moving steps of elements in row and column, respectively:
\begin{IEEEeqnarray}{rCl}
s = \frac{J-1}{2},t = \frac{K-1}{2},
\end{IEEEeqnarray},
where $J \times K$ is the size of the $PSF$ mask.

\end{lem}
\begin{pf}
The convolution structure or convolution meaning of a block circulant matrix has been discussed in the proof of Theorem 4.1, by which the proving of this lemma is natural.

The block-matrix multiplication of $H\alpha_1^s$ will make row convolutions satisfy the symmetric property. The multiplication of $C(\alpha_2^t)$ will result in multiplying each matrix of a block element by $\alpha_2^t$, such as
\begin{IEEEeqnarray}{rCl}
\pmatset{0}{4pt}
HC(\alpha_2^t)=\begin{pmat}({||})
H_0\alpha_2^t & H_{M - 1}\alpha_2^t  \cdots & H_1\alpha_2^t \cr\-
H_1\alpha_2^t & H_0\alpha_2^t \cdots & H_2\alpha_2^t \cr\-
\vdots & \ddots & \vdots \cr\-
H_{M-1}\alpha_2^t & H_{M-2}\alpha_2^t \cdots & H_0\alpha_2^t \cr
\end{pmat},
\end{IEEEeqnarray}
through which the symmetric property of column convolutions could be satisfied. The combination of operations of $\alpha_1^s$ and $C(\alpha_2^t)$ finally realizes the symmetric property of the two-dimensional $PSF$ convolution.
\end{pf}

\begin{thm}
In {\rm{(4-1)}}, if we change the PSF $h(m, n)$ into $h'(m, n)$ of {\rm(4-6)} and construct a convolution matrix $H_t'$ via $h'(m, n)$ by Lemma {\rm4.1}, and then
\begin{IEEEeqnarray}{rCl}
\vec{g} = H_t'\vec{f}
\end{IEEEeqnarray}
satisfies the symmetric property and the convolution structure is the original $h(m, n)$.
\end{thm}
\begin{pf}
The proof is the generalization of the one-dimensional case of Theorem 3.2. We only need to perform the operations satisfying the symmetric property in row and column dimensions by Lemma 4.1, as well as the reverse-order operations in two dimensions; the combined effects are in the form of (4-14).
\end{pf}

\begin{thm}
If the frequency domain formula {\rm{(4-2)}} is modified to
\begin{IEEEeqnarray}{rCl}
\mathcal{G}(u,v)= \mathcal{F}(u,v)\mathcal{H}'(u,v)e^{\frac{i2\pi}{N}vt}e^{\frac{i2\pi}{M}us},
\end{IEEEeqnarray}
where $\mathcal{H}'(u,v)$ is the two-dimensional $DFT$ of $h'_e(m, n)$ (the extended form of $h'(m, n)$ of {\rm(4-6))}, $s$ and $t$ are as define in {\rm(4-12)}, then its corresponding spatial domain convolution satisfies the symmetric property and the $PSF$ is the original $h(m, n)$.
\end{thm}
\begin{pf}
We decompose the proof into two kinds of one-dimensional case by the two-dimensional $DFT$ of $h_e(m, n)$:
\begin{IEEEeqnarray*}{rCl}
\mathcal{H}(u, v) = \sum_{m = 0}^{M-1}\sum_{n = 0}^{N - 1}h_e(m, n)e^{-i\frac{2\pi u}{M}m}e^{-i\frac{2\pi v}{N}n},
\end{IEEEeqnarray*}
which can be written as
\begin{IEEEeqnarray}{rCl}
\mathcal{H}(u, v) = \sum_{m = 0}^{M-1}h_e(m, n)e^{-i\frac{2\pi u}{M}m}\sum_{n = 0}^{N - 1}e^{-i\frac{2\pi v}{N}n}.
\end{IEEEeqnarray}

In (4-16), fixing $n$ and $v$, the left sum of the right side of (4-16) is the $DFT$ of row dimension of column $n$; so the right side of (4-16) can be considered as the sum of multiplications of a row-dimension $DFT$ multiplied by a factor $e^{-i\frac{2\pi v}{N}n}$; we can apply the one-dimensional result of Theorem 3.3 to the row dimension separately. After that, the column dimension can be dealt with. Combinations of the row and column yield the final result of (4-15).
\end{pf}

\begin{rmk-2}
As in the one-dimensional case, \rm{(4-14)} and \rm{(4-15)} are also corresponding to each other, which are the modified formulas in spatial domain and frequency domain, respectively.
\end{rmk-2}

\section{Applications of new formulas}
We mainly illustrate the effect of formula (4-15). In reality, the degraded image is produced physically by the $PSF$ convolution. It's noteworthy that the circular convolution in the margin of an image doesn't conform to the actual physical process of blur producing, due to the zigzag order or wrapping around operation; however, we obtain the blurred image by circular convolutions in this paper, only aiming to visualize the effect of the new formula (4-15).

\begin{figure}[!t]
\centering
\subfloat[]{\includegraphics[width=1.6in]{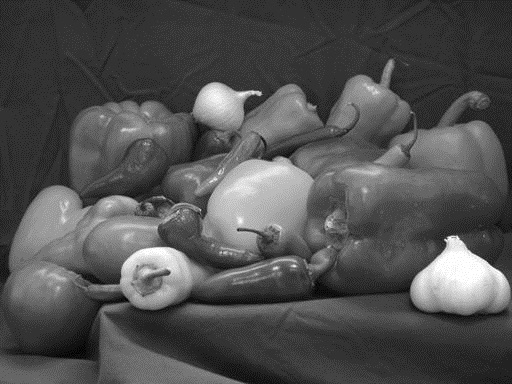}\label{a}} \quad
\subfloat[]{\includegraphics[width=1.6in]{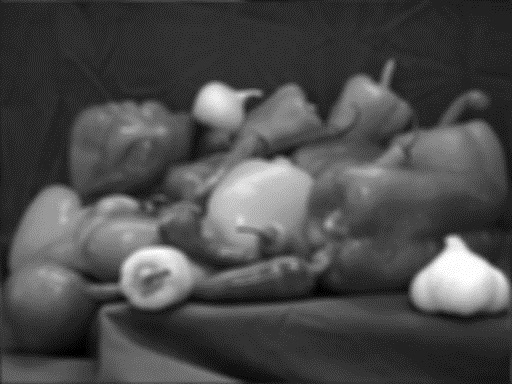}\label{b}} \\
\subfloat[]{\includegraphics[width=1.6in]{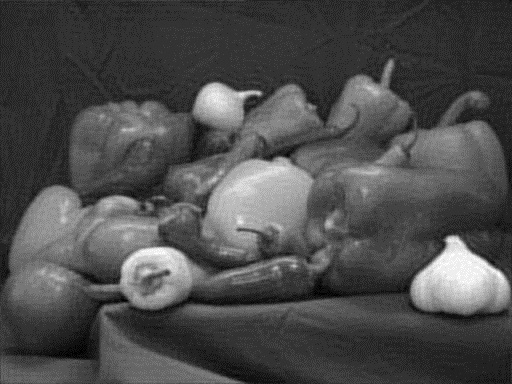}\label{c}} \quad
\subfloat[]{\includegraphics[width=1.6in]{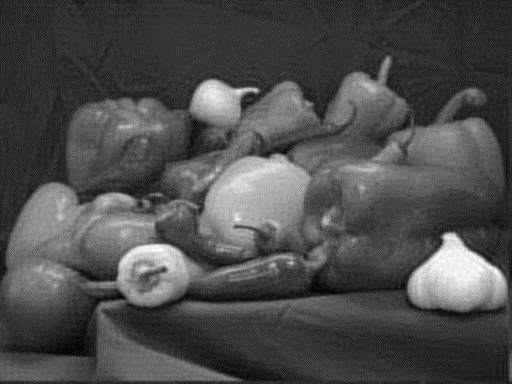}\label{d}} \\
\subfloat[]{\includegraphics[width=1.6in]{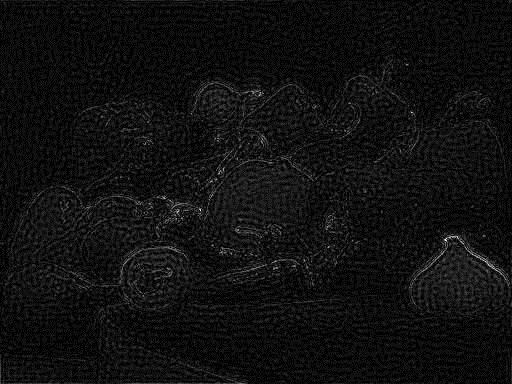}\label{e}} \quad
\subfloat[]{\includegraphics[width=1.6in]{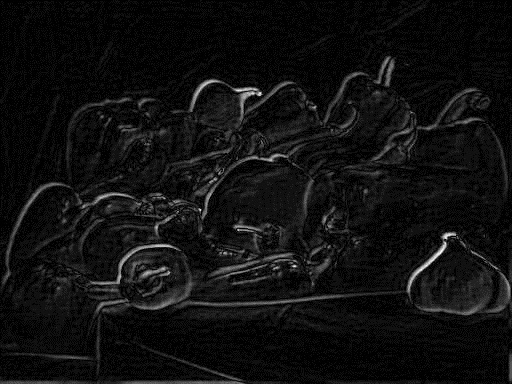}\label{f}} \\
\caption[]{Deconvolution example of uniform 2-D blur: \subref{a} Original image. \subref{b} Blurred image. \subref{c} Deblurred image by new formula (5-3). \subref{d} Deblurred image by formula (5-1). \subref{e} Normalized difference between deblurred image of (5-3) and original image. \subref{f} Normalized difference of (5-1).}
\label{Fig.6}
\end{figure}

Constrained least square estimation \cite{[5]} is a classical method of image restoration, whose frequency domain solution is
\begin{IEEEeqnarray}{rCl}
\mathcal{F}(u, v) = \frac{\overline{\mathcal{H}}(u, v)\mathcal{G}(u, v)}{\mathcal{H}(u, v)\overline{\mathcal{H}}(u, v) + \gamma\mathcal{C}(u, v)\overline{\mathcal{C}}(u, v)},
\end{IEEEeqnarray}
where $\overline{\mathcal{H}}$ is the complex conjugate of $\mathcal{H}$ and $\mathcal{C}(u, v)$ is the smooth regularization term. (5-1) can be expressed in the form of fundamental frequency domain formula
\begin{IEEEeqnarray}{rCl}
\mathcal{G}(u,v) = \mathcal{F}(u,v)\mathcal{H}(u,v) + \alpha\mathcal{F}(u,v)\mathcal{C}(u, v),
\end{IEEEeqnarray}
where $\alpha = \gamma\overline{\mathcal{C}}(u, v)/\overline{\mathcal{H}}(u, v).$

The smooth regularization is also operated by mask convolutions, so $\mathcal{C}(u, v)$ of (5-2) should be considered as the $DFT$ of another $PSF$. Both of $\mathcal{H}(u, v)$ and $\mathcal{C}(u, v)$ in (5-1) need to be modified by the result of (4-15). The modified version of (5-1) is
\begin{IEEEeqnarray}{rCl}
\mathcal{F}(u, v) = \frac{\overline{\mathcal{H}}_1(u, v)\mathcal{G}(u, v)}{\mathcal{H}_1(u, v)\overline{\mathcal{H}}_1(u, v) + \gamma\mathcal{C}_1(u, v)\overline{\mathcal{C}}_1(u, v)},
\end{IEEEeqnarray}
where
\begin{IEEEeqnarray*}{rCl}
\mathcal{H}_1(u, v) = \mathcal{H}'(u,v)e^{\frac{i2\pi}{N}vt}e^{\frac{i2\pi}{M}us}, \mathcal{C}_1(u, v) = \mathcal{C}'(u,v)e^{\frac{i2\pi}{N}vt}e^{\frac{i2\pi}{M}us}.
\end{IEEEeqnarray*}

Fig.\ref{Fig.6} is a deconvolution example of uniform 2-D blur \cite{[4]} images. Fig.\ref{Fig.6}(a) is an original image and Fig.\ref{Fig.6}(b) is a blurred image of Fig.\ref{Fig.6}(a), which is obtained by circular convolutions of a $PSF$ mask of size $7 \times 7$. Fig.\ref{Fig.6}(c) is the deblurred image by new formula (5-3) and Fig.\ref{Fig.6}(d) is the result of formula (5-1). Fig.\ref{Fig.6}(c) and Fig.\ref{Fig.6}(d) seem to have no differences apparently, but when comparing the normalized difference between the deblurred image and the original image, as shown in Fig.\ref{Fig.6}(e) and Fig.\ref{Fig.6}(f), it's obvious that Fig.\ref{Fig.6}(c) is much better than Fig.\ref{Fig.6}(d), since its normalized difference is more uniform and much closer to zero.

\section{Conclusions}
The symmetric property of the $PSF$ convolution is the intrinsic physical property of the producing of defocus blur and atmospheric-turbulence blur. Although motion blur doesn't belong to that case, in mathematical form of the $PSF$ mask model, it can be treated in the same way, which is necessarily required in blind deconvolution of motion blur.

Circular convolutions and the corresponding frequency domain formula are widely used in image processing; however, the usual computing method violates the physical meaning of blur producing. We proved this problem rigorously and gave the modified formulas.

For clarity of proving, we also dealt with the one-dimensional case, whose results may be useful in one-dimensional signal processing.





\end{document}